\documentclass[conference]{IEEEtran}
\IEEEoverridecommandlockouts
\usepackage{amsmath, amssymb, amsfonts}
\usepackage{graphicx}
\usepackage{xcolor}
\usepackage{wrapfig}
\usepackage{comment}
\usepackage{textcase}
\usepackage{authblk}
\usepackage{subcaption}

\newcommand*{\affmark}[1][*]{\textsuperscript{#1}}
\usepackage{multicol}
\usepackage{color}
\usepackage{graphicx}
\usepackage{soul}
\usepackage{epstopdf}
\usepackage{physics}

\def\shahab{\textcolor{magenta}}

\usepackage{comment}
\usepackage[numbers, sort&compress, square]{natbib}
\usepackage{enumerate}
\usepackage{mathtools}
\usepackage{amsmath}
\usepackage{bm}
\usepackage{newtxtext, newtxmath}
\usepackage{adjustbox}
\usepackage[noend]{algpseudocode}
\usepackage{pdfpages}
\usepackage[font=small,skip=0pt]{caption}

\usepackage[normalem]{ulem}
\usepackage{float}
\usepackage{scalerel}
\usepackage{tikz}
\usepackage{breqn}
\usepackage{algpseudocode}
\usetikzlibrary{svg.path}

\usepackage[ruled,vlined ]{algorithm2e}
\definecolor{orcidlogocol}{HTML}{A6CE39}
\tikzset{
orcidlogo/.pic={
\fill[orcidlogocol] svg{M256, 128c0, 70.7-57.3, 128-128, 128C57.3, 256, 0, 198.7, 0, 128C0, 57.3, 57.3, 0, 128, 0C198.7, 0, 256, 57.3, 256, 128z};
\fill[white] svg{M86.3, 186.2H70.9V79.1h15.4v48.4V186.2z}
         svg{M108.9, 79.1h41.6c39.6, 0, 57, 28.3, 57, 53.6c0, 27.5-21.5, 53.6-56.8, 53.6h-41.8V79.1z M124.3, 172.4h24.5c34.9, 0, 42.9-26.5, 42.9-39.7c0-21.5-13.7-39.7-43.7-39.7h-23.7V172.4z}
         svg{M88.7, 56.8c0, 5.5-4.5, 10.1-10.1, 10.1c-5.6, 0-10.1-4.6-10.1-10.1c0-5.6, 4.5-10.1, 10.1-10.1C84.2, 46.7, 88.7, 51.3, 88.7, 56.8z};
}
}

\newcommand\orcidicon[1]{\href{https://orcid.org/#1}{\mbox{\scalerel*{
\begin{tikzpicture}[yscale=-1, transform shape]
\pic{orcidlogo};
\end{tikzpicture}
}{|}}}}

\def\BibTeX{{\rm B\kern-.05em{\sc i\kern-.025em b}\kern-.08em
T\kern-.1667em\lower.7ex\hbox{E}\kern-.125emX}}

\SetKwInput{KwInput}{Input}                
\SetKwInput{KwOutput}{Initialization}      

\begin{document}

\title{ Distributed Dynamic Economic Dispatch using Alternating Direction Method of Multipliers}

\author{
Shailesh Wasti\affmark[1], Pablo Ubiratan\affmark[1], Shahab Afshar\affmark[1] and Vahid Disfani\affmark[1] \\ \affmark[1]{ConnectSmart Research Laboratory, University of Tennessee at Chattanooga, TN 37403, USA}\\ 
emails: shailesh-wasti@mocs.utc.edu,  vahid-disfani@utc.edu
}

\maketitle

\begin{abstract}
With the proliferation of distributed energy resources and the volume of data stored due to advancement in metering infrastructure, energy management in power system operation needs distributed computing. In this paper, we propose a fully distributed Alternating Direction Method of Multipliers (ADMM) algorithm to solve the distributed economic dispatch (ED) problem, where the optimization problem is fully decomposed between participating agents. In our proposed framework, each agent estimates the dual variable and the average of the total power mismatch of the network using dynamic average consensus, which replaces the dual updater in the traditional ADMM with a distributed alternative. Unlike other distributed ADMM, the proposed method does not rely on any specific assumption and captures the real-time demand change. The algorithm is validated successfully via case studies for IEEE 30-bus and 300-bus test systems with the penetration of solar photovoltaic.
\end{abstract}


\section{Introduction}
The steady growth of distributed energy resources (DER), due to the society’s increasing commitment to a low-carbon economy along with the remarkable progress in technology, is questioning the relative benefit of large-scale electricity generation that is supposed to stem from high economies of scale. The disruption at the grid edge of the physical layers of the power system--generation, transmission, and distribution--has profoundly transformed the 21st-century power systems, and the time has come to revisit the classic categorization of these physical layers. The explosion of data with the proliferation of DER has allowed researchers to look for new ways to control, monitor, and optimize resources. The dispatch of generators based on their marginal cost of production--known as economic dispatch (ED)--has become challenging because of the number of players in the electricity markets and their hesitation to share their own control variables and economic data due to privacy and economic reasons. 
To address these concerns, multi-agent algorithms are projected as promising solutions to distribute the economic dispatch problem between agents. 

There are a significant number of research works in the literature on decentralized and distributed methods including analytical target cascading (ATC) \cite{kargarian2016toward} and alternating direction method of multipliers (ADMM) \cite{boyd2011distributed}. ADMM has been projected as a robust and powerful method to decouple the centralized optimization problem. ADMM with model predictive control is used to control and schedule DER in real-time in a microgrid \cite{wang2015dynamic}, but with a need of a central coordinator to push the decision variable and clear the market. The central idea in this method is that the central coordinator gathers the optimal decision variables from all the agents to compute the global dual variable and distributes the necessary information back to the agents. In economic dispatch problems and many other optimization problems on resource allocation, this dual variable is the market-clearing price that is updated to eliminate the power mismatch of the network. However, in these ADMM frameworks, the need for central coordinator to communicate directly with all agents involved may still cause privacy concerns for the private agents.

To address the privacy concerns, distributed platforms are considered as state-of-the-art solutions where each agent just communicates with a limited number of agents as neighbors, and they update the dual variables themselves \cite{kargarian2016toward}. To design distributed algorithms, all global variables must be localized and some consensus algorithms must be employed. In order to localize the global variables, there are several solutions proposed including primal-dual \cite{disfani2015distributed} and center-free algorithms \cite{xu2013distributed}. The concept is to make agents reach consensus on global variables independently, without a need of a central coordinator. \cite{xu2013distributed} uses frequency as a proxy for power balance constraint to coordinate among renewable generators in an attempt to achieve a fully distributed platform, but at the expense of the power system model (central coordinator) for the local frequency measurement. 

In \cite{chen2017admm}, a distributed ADMM platform is proposed for economic dispatch problem without a need for a central coordinator or a leader. An average consensus algorithm and projection methods are used to find the consensus of the primal variables, \textit{i.e.} the agents' power generation. Despite of the effectiveness of the algorithm, it suffers from a strong assumption that requires the initial condition to start from a zero-power-mismatch condition. That is total generation and total demand must be equal at the first iteration for the algorithm to work. The algorithm also cannot capture the dynamics of demand change, which makes it impractical for real-time implementations with high penetration of variable renewable energy resources.

In this paper, a fully distributed ADMM algorithm is proposed to solve the real-time economic dispatch problem. There are three main contributions associated with the proposed algorithm, none of which has been addressed in the literature before.
\begin{enumerate}
\item	The need for dual updater in the ADMM algorithm is removed by exploiting \textit{dynamic average consensus} algorithm reported in \cite{spanos2005dynamic, spanos2005distributed}. In our proposed algorithm, each agent solves its optimization problem for the given market price and their own estimates of the dual variable and the average of the total power mismatch of the network by communicating with their neighbors only. This makes the ADMM algorithm fully distributed which is applied to the distributed economic dispatch problem.
\item The agents' privacy is preserved better in our algorithm as agents communicate the estimate of the average of the total power mismatch to their neighbors instead of communicating the generated power and the demand at their node and carries little information about agents' private information. 
\item The proposed algorithm is robust enough to capture the dynamic demand changes, which is more significant than before with increasing penetration of variable renewable energy in power grids.
\end{enumerate}

The algorithm is tested against two IEEE test systems at different scales (30-bus and 300-bus) with high PV penetrations. The results demonstrates that the proposed algorithm is robust and scalable enough for real-time applications.  

The remainder of the paper is presented as follows. Section~\ref{sec:GT n consensus Maths} provides mathematical background on graph theory and dynamic consensus algorithm. section \ref{sec:D-ED} elaborates the design of the distributed economic dispatch algorithm. Section \ref{sec:simulation} presents and discusses the simulation results. Section \ref{sec:conclusion} concludes the paper. 

\section{Preliminaries on Graph Theory and Dynamic Average Consensus Algorithm} 
\label{sec:GT n consensus Maths}
There are many excellent texts on graph theory and linear iteration. We refer our discussion  from \cite{newman2018networks} and restrict it to an \textit{undirected, simple} graph. Matrices and vectors are written in bold throughout the paper. 
\subsection{Notation and Graph Theory}
Let ${\mathcal{G}}$ denote a graph with a set of vertices $\mathcal{V}$ and a set of edges $\mathcal{E}$. Let ${v_i} \in \mathcal{V}$ for all  $i \in [1, N]$ be the node  and $e=(v_i, v_j)$ $\forall i$ be the edges of the graph, where $v_i$ is referred as head and $v_j$  as tail. Our network has neither self-edges nor multi-edges,  or simply $v_i \neq v_j~\forall i, j$. The cardinality of node, $\left | V \right |$ is N and that of edge, $\left | E \right |$ is M. Let $d_i$ is the degree of node $v_i$ defined as the total number of its  neighbors. 

Let $\textbf{A}$ be the \textit{adjacency matrix} of graph ${\mathcal{G}}$, and {\textbf{D}} be the degree matrix with the vertex degree along its diagonal. Then, the  \textit{graph Laplacian} $\mathbf{L}\in\mathbb{R}^{N\times N}$ can be expressed in matrix form as:
\begin{align}
\textbf{L}=\textbf{D}-\textbf{A}
\end{align}
In full form:
\begin{align}
l_{ij}:=\left\{\begin{array}{ll}{\operatorname{d_i}} &  {i=j,} \\ {-1} & {i \neq j \text{and there is an edge (i,j)}} \\ {0} & {\text{otherwise}}\end{array}\right.
\end{align}

The \textit{Laplaican} consensus dynamics is given by the equation  \cite{spanos2005dynamic}
\begin{align}
\textbf{$\dot{x}$}=-\textbf{Lx}
\end{align}

The following propositions can be made about the matrix \textbf{L}.
\begin{enumerate}
\item {\textit{Laplacian} is a symmetric matrix, and has real eigen values.}
\item{All eigen values of the \textit{Laplacian} are non-negative.}
\item{\textit{Laplacian} always has at least one zero eigen values, that is $\mathbf{L. 1 =0 }$}
\item {The second smallest eigen value of \textit{Laplacian graph} is non-zero if and only if the network is connected, and hence the value is called algebraic connectivity of the network}
\end{enumerate}

The properties of \textbf{L} show that the dynamics is stable and convergent, which take any random initial value to consensus. It can be summarized in the following equation:
\begin{align}
\frac{1}{N} \mathbf{1}^T \textbf{x}_0 =\frac{1}{N} \sum_i {x}_{i}^{0}
\end{align}
\subsection{Dynamic Average Consensus}
\label{Sec:Avg_COn}
Let subscript $i$ denote the initial value {$\mathbf{{x}_{i}^{0}\in \mathbb{R}^{N}}$}, and  $x(0)=(x_1(0), \cdots , x_N(0))$ be the vector of the initial values of the network. The discussion below is on the computation of average, $ \frac{1}{N} \sum_{i=1}^{N} {x}_{i}^{0}$, in distributed fashion, where each node in the graph communicates with only its neighbor.
Let us consider \textit{distributed linear iteration} of the form
\begin{align}
x_{i}^{k+1}=a_{i i} x_{i}^{k}+\sum_{j \in \mathcal{N}_i} a_{i j} x_{j}^{k}, \quad i=1, \ldots, N
\end{align}
where $ k=0, 1, 2, \ldots$ and $ a_{i j} $ is the weight on $x_j$ at node $i$, and $\mathcal{N}_i$ denotes the set of all neighbors of agent $i$. Setting $a_{i j}=0$ for $j \notin \mathcal{N}_{i}$, this iteration can be written as 
\begin{align}
\mathbf{x}_{}^{k+1} = \mathbf{Ax}_{}^{k}
\label{eq:avg_consensus}
\end{align}
According to \cite{xiao2004fast}, the constraint on the sparsity pattern of the matrix $ \bf{A} $ can be expressed as $\bf{A} \in \mathcal{S}$, where  \begin{align}
\mathcal{S}=\left\{A \in \mathbf{R}^{N \times N} | A_{i j}=0 \text { if }\{i, j\} \notin \mathcal{E} \text { and } i \neq j\right\}
\end{align}
and \eqref{eq:avg_consensus} can be written as 
\begin{align}
\mathbf{x}_{}^{k} = \mathbf{A^k x}_{i}^{0}, \quad k =0, 1, 2, \cdots 
\label{eq:iteration_b}
\end{align} 
We are interested in the matrix ${\mathbf{A}}$ such that:
\begin{align}
    {\mathbf{A}}^k=\frac{\mathbf{11}^T}{N} \quad \text{as}\quad k\rightarrow\infty
    \label{eq:cgt_Matrix}
\end{align}
 As it is  shown in \cite{xiao2004fast}, the left hand side of \eqref{eq:cgt_Matrix} converges if and only if:
 \begin{align}
 \label{Eq:necessary cond}
   \mathbf{1}^T \mathbf{A}=\mathbf{1}^T \\
   \mathbf{A 1}=\mathbf{1}\\
   \rho(\mathbf{A}-\mathbf{11}^T/N)<1
   \end{align}
  where {$\rho(\cdot)$} is the spectral radius of the matrix.

There are different ways to set the coefficients of matrix $\mathbf{A}$ such as \textit{constant edge Weight} and \textit{local degree weight}. We are interested to assign weight that only depends on the local information. The  \textit{local degree weight} matrix $\mathbf{A}$ with coefficients that depend only the degree of the incident node is:    
\begin{align}
{a}_{i j}=\left\{\begin{array}{cl}\frac{1}{\text{max}\{d_i,d_j\}} & {\{i, j\} \in \mathcal{E}} \\ 1-\sum_{j\in \mathcal{E}}\frac{1}{\text{max}\{d_i,d_j\}} & {i=j} \\ {0} & {\text { otherwise }}\end{array}\right.
\label{eq:metro}
\end{align}

 This method comes from the Metropolis-Hastings algorithm and often called the \textit{Metropolis} method.
 The improved \textit{Metropolis} called \textit{Mean Metropolis} is proposed in \cite{xu2011novel} where 

\begin{align}
{a}_{i j}=\left\{\begin{array}{cl}\frac{2}{\{d_i+d_j+\epsilon \}} & {\{i, j\} \in \mathcal{E}} \\ 1-\sum_{j\in \mathcal{E}}\frac{2}{\{d_i+d_j+\epsilon\}} & {i=j} \\ {0} & {\text { otherwise }}\end{array}\right.
\label{eq:mean met}
\end{align}
where $\epsilon$ is a very small number.

The average consensus described by \eqref{eq:avg_consensus} can be extended to the network in which the signal in each node is the function of time \cite{spanos2005dynamic}.

Let $\Delta \mathbf{z}$ be the input bias applied to \textit{average consensus system}.

We can claim that the following modification to \eqref{eq:avg_consensus} makes the \textit{dynamic consensus} algorithm tracks the time-varying \textit{average consensus}:
\begin{align}
\mathbf{x}_{}^{k+1} = \mathbf{A x}_{}^{k} +\Delta \mathbf{z}
\label{eq:Dynamic consensus}
\end{align}
where the bias $\Delta \mathbf{z} = \mathbf{z}^{k+1}-\mathbf{z}^{k}$.

The modification is still distributed as each agent has their own bias and the algorithm is convergent as it still has the conservation property proved in \ref{sec:Appndx-Cosv prop}. 
\section{Distributed Economic Dispatch}
\label{sec:D-ED}
Economic Dispatch is the determination of the optimal output of each generator based on the marginal cost of production in order to meet the demand of the network.
Economic Dispatch is defined as follows:

{\emph{Economic Dispatch (ED):}}
\begin{subequations}
\allowdisplaybreaks
\begin{align}
\min_{{P_{g}}} & \sum_{i\in \mathcal{V}} C_i(P_{gi}) \label{P2:obj}\\
\text{s. t.} & \sum_{i\in \mathcal{V}} P_{gi} = \sum_{i\in \mathcal{V}} P_{di}\label{P2:totalpowerbalance}\\
& {P_{gi}^{\min}}\le P_{gi} \le P_{gi}^{\max} &\forall i\in \mathcal{V}\label{P2:genlimit}
\end{align}
\label{P2}
\end{subequations}

Note that \eqref{P2:totalpowerbalance} is the only constraint that requires power balance across the entire network. If (\ref{P2:obj}) and (\ref{P2:genlimit}) are convex, the optimization problem (\ref{P2}) is a convex problem. 

\subsection{ADMM for Economic Dispatch}
\label{Sec:ADMM for ED}
ADMM is an extension of the method of multipliers but can be decomposed into sub-problems. The scaled form of it can be summarized in the following equations \cite{boyd2011distributed}: 

\begin{equation}
\begin{split}
P_{gi}^{k+1}= {} & \operatorname*{arg min}_{p^{min}_{gi}\leq p_{gi} \leq p^{max}_{gi}} \Bigg\{ c_i(P_{gi})\\ & +\frac{\rho }{2} \Bigg\|{\sum_j P_{dj}^k-(P_{gi} 
+ \sum_{j \neq i} P_{gj}^k)+ U^k} \Bigg\|_2^2  \Bigg\} 
\label{eq:Quasi_ADMM_1}
\end{split}
\end{equation}
\begin{align}
  U^{k+1}=U^{k}+\sum P_{di}^{k+1}-\sum P_{gi}^{k+1}
  \label{eq:dual_1}
\end{align}
where $\rho > 0$ is the penalty parameter, $\|\cdot\|_2$ is the usual Euclidean norm, and $U$ is the dual updater. 

At each iteration, the dual updater which updates $U$ gathers the generation and the demand information from all the agents to calculate the supply-demand imbalance of the network. The dual variable is updated to drive the residue from the network's power mismatch to zero. In other words, $U$ in \eqref{eq:dual_1} is constantly pulling the decision variables toward the optimal value projected onto the feasible space. In terms of communication, each agent receives the updated dual variable and the generation and the demand information of all agents but itself, which sums to $2(N-1)+1$ numbers of data. 

Let $\overline{P}_{gd}$ denote the average power mismatch between generation and demand given as
\begin{align*}
 \overline{P}_{gd} = \frac{1}{N} \left(\sum_{i=1}^N P_{gi}-\sum_{i=1}^N P_{di}\right). 
\end{align*}

Then, \eqref{eq:Quasi_ADMM_1}-\eqref{eq:dual_1} can be written as 
\begin{equation}
\begin{split}
P_{gi}^{k+1}={} & \operatorname*{arg min}_{p^{min}_{gi}\leq p_{gi} \leq p^{max}_{gi}} \Bigg\{ c_i(P_{gi})\\ & + \rho/2 \left\| {P_{gi}^k-N\overline{P}_{gd}^{k}-P_{gi}+U^k}\right\|_2^2 \Bigg\} 
\label{eq:Quasi_ADMM_2}
\end{split}
\end{equation}
\begin{align}
      U^{k+1}=U^{k}-N \overline{P}_{gd}^{k+1}
  \label{eq:dual_2}
\end{align}

Denoting $W=\frac{U}{N}$, \eqref{eq:Quasi_ADMM_2}-\eqref{eq:dual_2} can be simplified as:
\begin{equation}
\begin{split}
P_{gi}^{k+1}= {} & \operatorname*{arg min}_{p^{min}_{gi}\leq p_{gi} \leq p^{max}_{gi}} \Bigg\{ c_i(P_{gi})\\ & +\frac{\rho N^2}{2} \Bigg\| -\frac{P_{gi}}{N}+ {\frac{{P^k}_{gi}}{N}-\overline{P}_{gd}^k+W^k}\Bigg\|_2^2  \Bigg\} 
\label{eq:Quasi_ADMM_3}
\end{split}
\end{equation}
\begin{align}
  {W}^{k+1}={W}^k-\overline{P}_{gd}^{k+1}  \label{eq:dual_3}
\end{align}

 Unlike in the formulation \eqref{eq:Quasi_ADMM_1}-\eqref{eq:dual_1} which needs large bandwidth for communication, the dual updater in \eqref{eq:Quasi_ADMM_3}-\eqref{eq:dual_3} needs to communicate only three variables namely average power mismatch of the network, updated dual variable, and number of agents. The modification in \eqref{eq:Quasi_ADMM_3}-\eqref{eq:dual_3}, thus, reduces the communication burden in the sub-problems, except that it is still not fully distributed. 

\subsection{Distributed-ADMM for Economic Dispatch}
We leverage the concept of dynamic average consensus discussed in \ref{Sec:Avg_COn} to convert all the three global variables $\overline{P}_{gd}$, $U$, and $N$ into local variables. We adopt the graph discovery algorithm method proposed in \cite{guo2015distributed} to localize $N$. Localization of the other two parameters are proposed below.
\subsubsection{Dynamic Power mismatch Consensus Algorithm}
In order to make the algorithm in \ref{Sec:ADMM for ED} fully distributed, every agent needs to calculate the average generation and the average demand of the network which is changing in time, and the effect is more pronounced with the increasing penetration of unpredictable and variable renewable energy resources including solar photovoltaic systems. With reference to \eqref{eq:Dynamic consensus}, the local variables of $\overline{P}_{gi}^k$, $\overline{P}_{di}^k$, and $\overline{P}_{gdi}^k$ which are defined as agent $i$'s estimate of the average generation, demand, and power mismatch of the entire network at iteration $k$, can be calculated as 
\begin{align}
&\overline{P}_{gi}^{k+1}=a_{i i} \overline{P}_{gi}^{k}+\sum_{j \in \mathcal{N}_i} a_{i j} \overline{P}_{gj}^{k}+(P_{gi}^{k+1}-P_{gi}^k) \label{eq:Pg_mean} \\
&\overline{P}_{di}^{k+1}=a_{i i} \overline{P}_{di}^{k}+\sum_{j \in \mathcal{N}_i} a_{i j} \overline{P}_{dj}^{k}+({P}_{di}^{k+1}-{P}_{di}^{k})\label{eq:Pd_mean}
\end{align}
Subtracting \eqref{eq:Pd_mean} from \eqref{eq:Pg_mean} and substituting ${P_{gdi}}$ for ${P_{gi}}-{P_{di}}$:
\begin{align}
    \overline{P}_{gdi}^{k+1}=a_{i i}\overline{P}_{gdi}^{k}+\sum_{j \in \mathcal{N}_i} a_{i j} \overline{P}_{gdj}^{k}+({P}_{gdi}^{k+1}-{P}_{gdi}^{k}) \label{eq:Pgd estimate}
\end{align}

Using \eqref{eq:Pgd estimate}, every agent can find consensus on average power mismatch of the network based on its own information and its neighbours'. This not only ensures that $\overline{P}_{gdi}$ is distributed, but also preserves privacy of the agents. It is almost impossible for an external agent to track the private information like the characteristic parameters of the generations such as cost function.   

\subsubsection{Dual Variable Consensus Algorithm}
With \eqref{eq:Pgd estimate}, the local dual variable in \eqref{eq:dual_3} is calculated in a distributed fashion as    
\begin{align}
  W_i^{k+1}=a_{i i} {W_i}^k+\sum_{j \in \mathcal{N}_i} a_{i j} {W_i}^k-\overline{P}_{gdi}^{k+1}  \label{eq:dual consensus}
\end{align}
The dual variable $W_i$ is a scaled variable of the market price $\lambda_i$ defined as:
\begin{align}
    \lambda_i^{k+1}=\rho N {W_i}^k \label{eq:market price}
\end{align}
With these formulations where all variables are calculated based on the agent $i$'s own data or its neighbors', a fully distributed ADMM equation boils down to:
\begin{equation}
\begin{split}
P_{gi}^{k+1}= {} & \operatorname*{arg min}_{p^{min}_{gi}\leq p_{gi} \leq p^{max}_{gi}} \Bigg\{ c_i(P_{gi})\\ & +\frac{\rho N^2}{2} \Bigg\| -\frac{P_{gi}}{N}+ {\frac{{P^k}_{gi}}{N}-\overline{P}_{gdi}^k+Wi^k}\Bigg\|_2^2  \Bigg\} 
\label{eq:ADMM:Pg*} 
\end{split}
\end{equation}
If the cost function of any generator $i$ is modeled as a quadratic function given by 
\begin{align}
C_{i}(P_{gi}) =  a_{i} P_{gi}^{2} + b_{i} P_{gi} + c_{i},
\end{align}
differentiating \eqref{eq:ADMM:Pg*} with respect to $P_{gi}$ leads to the optimal decision of 
\begin{align} 
\allowdisplaybreaks
P_{gi}^{k+1}:=\left[\frac{N\rho \uppsi_i^k -b_i}{2a_i+\rho}\right]_{{P_{gi}}^{min}}^{{P_{gi}}^{max}} \label{Pgi_opt}
\end{align}
where ${\uppsi_i^k=\frac{{P_{gi}}^k}{N}-\overline{P}_{gdi}^k+W_i^k}$. \textbf{Algorithm \ref{Alg:ADMM}} elaborates the entire update processes for the distributed ADMM method proposed for economic dispatch problem.

\begin{algorithm}
\KwInput{$\rho$}
  \KwOutput{{${W^0}_i,P^0_{gi},\overline{P}_{ gdi}^0$}}
        \For{$t=0$ to $\infty$}    
        { 
             calculate $P_{gi}^{k+1}$ using \eqref{eq:ADMM:Pg*}\\
             update $\overline{P}_{gdi}$ using \eqref{eq:Pgd estimate} \\
             update $ W_i$ using \eqref{eq:dual consensus}\\
             update $\lambda_i$ using \eqref{eq:market price}
        }
\caption{Executed by each agent $i \in \mathcal{V}$}
\label{Alg:ADMM}
\end{algorithm}

\section{Results and Discussions} \label{sec:simulation}
\subsection{Simulation Setup}
\label{Sec:Simulation_Setup}
In order to introduce variable renewable generation, we simulate the case where the installed capacity of roof-top solar equals to the demand at the node as specified in IEEE benchmark. We normalize the real data of the irradiance  starting from 3 PM April 2018 in 10 second resolution taken from from the PV Power Research Plant of Tampere University \cite{Jussi:2018}. The normalized irradiance $\widehat{Ir}$ is defined as the ratio of the irradiance at the time to the maximum irradiance observed in the month of April. Let $\Tilde{P}_d$ be the demand from the IEEE test cases, then the demand in each node $i$ is:
\begin{align*}
    {P_{di}}=\Tilde{P}_{di}-\widehat{Ir}_i P_{PV}
\end{align*}

We test the algorithm \ref{Alg:ADMM} against IEEE 30-bus and 300-bus test cases in $MATLAB R2019b$ environment. The other simulation parameters are provided in Table  \ref{tab:gen_param}. 
\begin{table}[!htb]
\centering
\caption{Case Study Parameters} 
\begin{tabular}{|c|c|}
\hline
 \textbf{Parameters} & \textbf{Values}\\
 \hline
\# of iterations ($k$) & 1e5  \\
 \hline 
 $\bf{\overline{P}_g(0)}$ & \bf{0} p.u. \\
\hline
  $\bf{\overline{P}_d(0)}$ & $\mathbf{P_d(0)}$ \\
\hline
 $\lambda(0)$ & $\sim U(MC_{min},MC_{max}) $ \\
 \hline
 Demand change step & every 20k iterations \\ 
\hline
 $N^2 \rho$ & 0.063546 \\
 \hline
 $\epsilon$ & $1$ \\
 \hline
 $a_{ij}$ & using \eqref{eq:mean met} \\
 \hline
 PV Installed capacity($P_{PV}$)  & 100\% of individual load \\
 \hline
\end{tabular} 
\label{tab:gen_param}
\end{table}

\subsection{Simulation Results}
\begin{table*}[h!]

\centering
\caption{$\lambda$ convergence table } 
\begin{tabular}{|c|c|c|c|c|c|c|}
\hline
{Test} & {Total} & {Total} & \multicolumn{2}{c|}{Initial Convergence} & \multicolumn{2}{c|}{Subsequent Convergence}\\
\cline{4-7} {Case} & {Time} & {Iteration} & \multicolumn{1}{l|}{\# iteration} & \multicolumn{1}{l|}{Time [sec]} & \multicolumn{1}{l|}{\# iteration} & \multicolumn{1}{l|}{Time [sec]} \\ \hline
Case 30  & 1.9386 & 1e5 & 400 & 0.007 & 300 & 0.005\\ \hline
Case 300 & 7.1194 & 1e5 & 10000 & 0.711 & 5000 & 0.355                           \\ \hline
\end{tabular}
\label{table: lambda convergence}

\end{table*}

\subsubsection{Test Cases}
\begin{figure}[h!]
    \begin{minipage}{1\linewidth}
        \begin{minipage}{1\linewidth}
            \includegraphics[width=\linewidth]{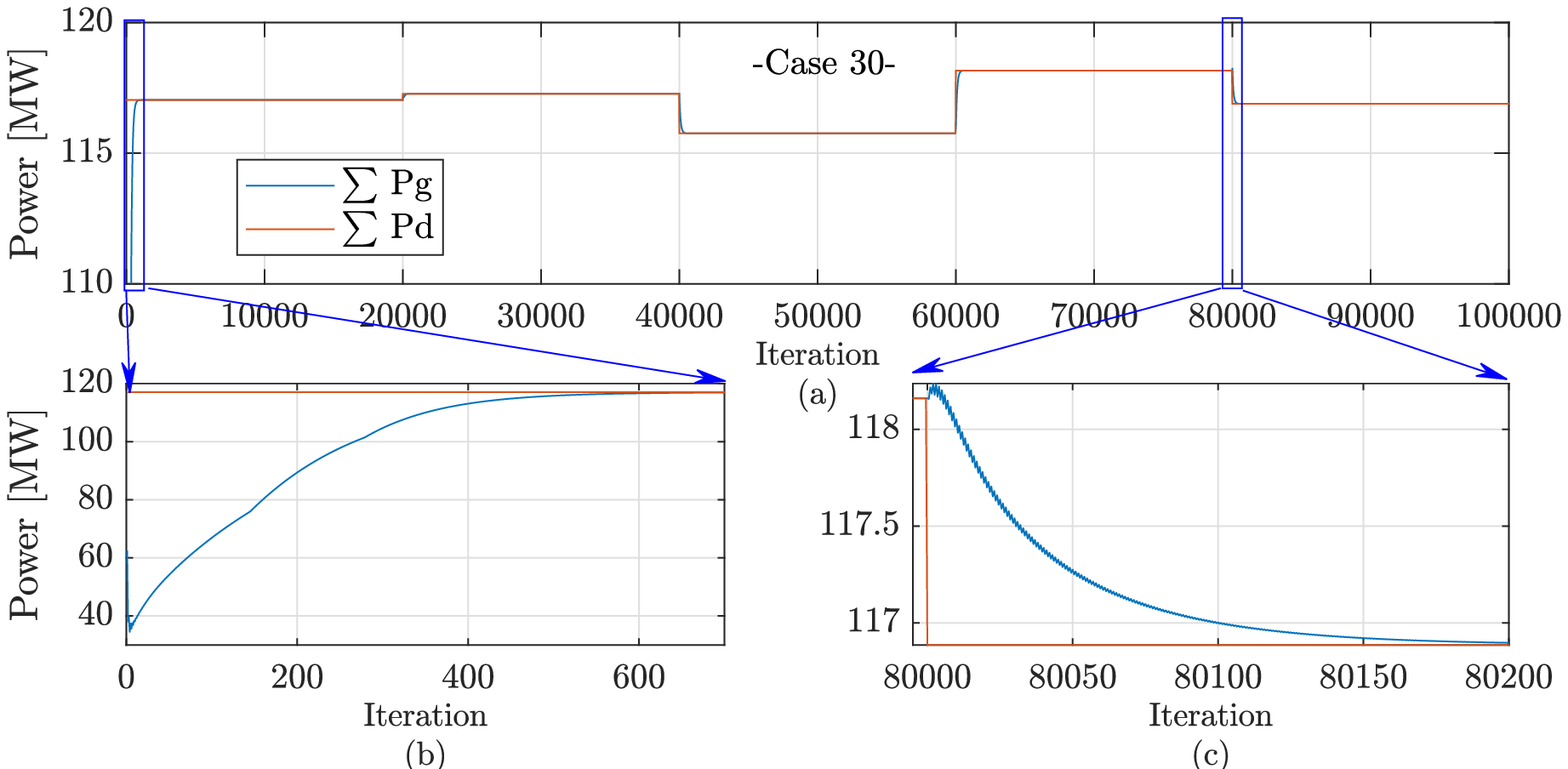}
        \end{minipage}
        \vfill
        \begin{minipage}{1\linewidth}
            \includegraphics[width=\linewidth]{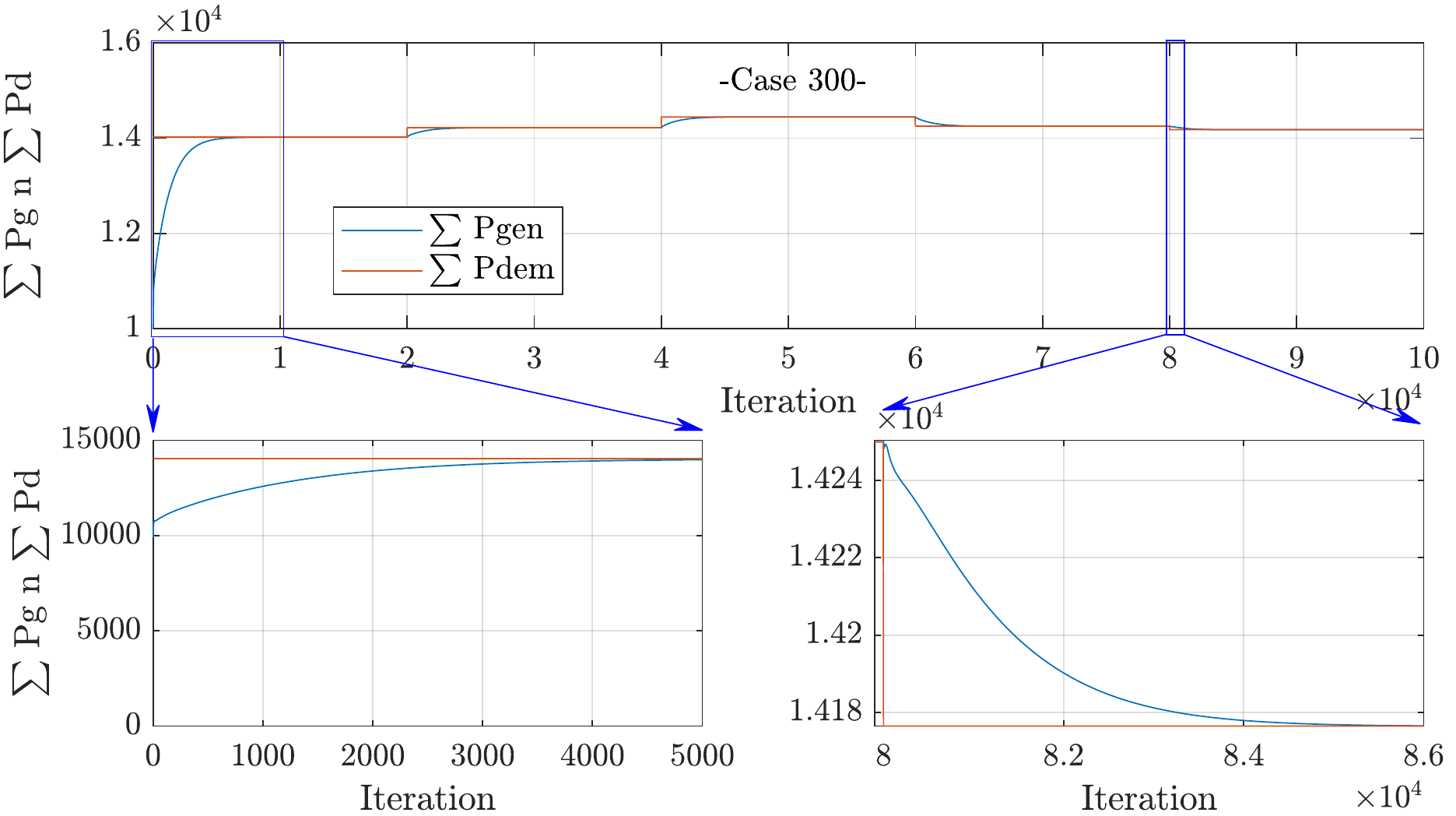}
        \end{minipage}
    \end{minipage}
    \caption{(a) and (d):Network's generation following demand  for IEEE 30-bus and 300-bus respectively;(b) and (e) are magnified version at the beginning , and (c) and (f) in the middle when the demand changed for 30-bus case and 300-bus respectively}
    \label{fig:300bus_Pgen_Pdem}
\end{figure}
\begin{figure}[h!]
    \begin{minipage}{1\linewidth}
        \begin{minipage}{1\linewidth}
            \includegraphics[width=\linewidth]{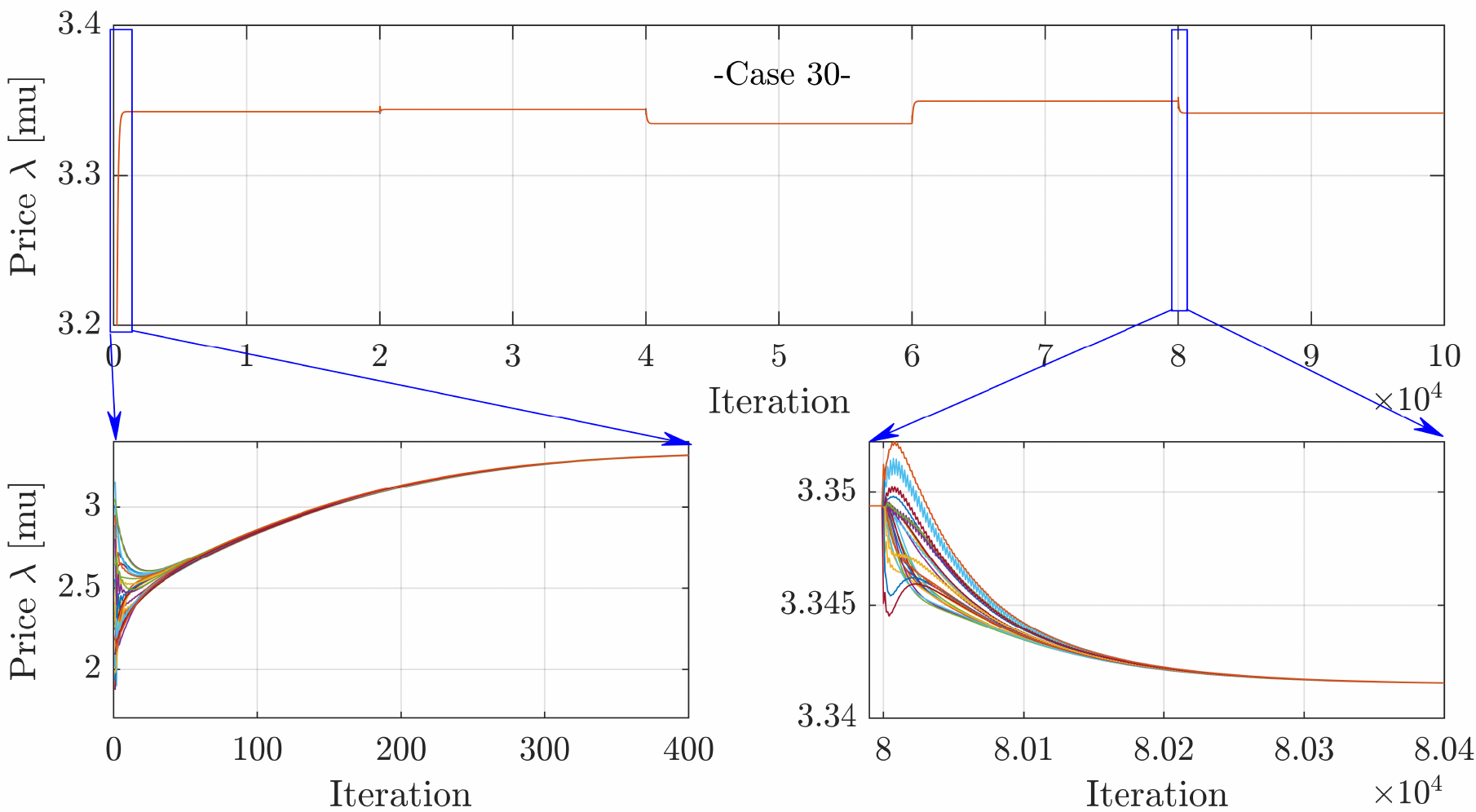}
        \end{minipage}
        \vfill
        \begin{minipage}{1\linewidth}
            \includegraphics[width=\linewidth]{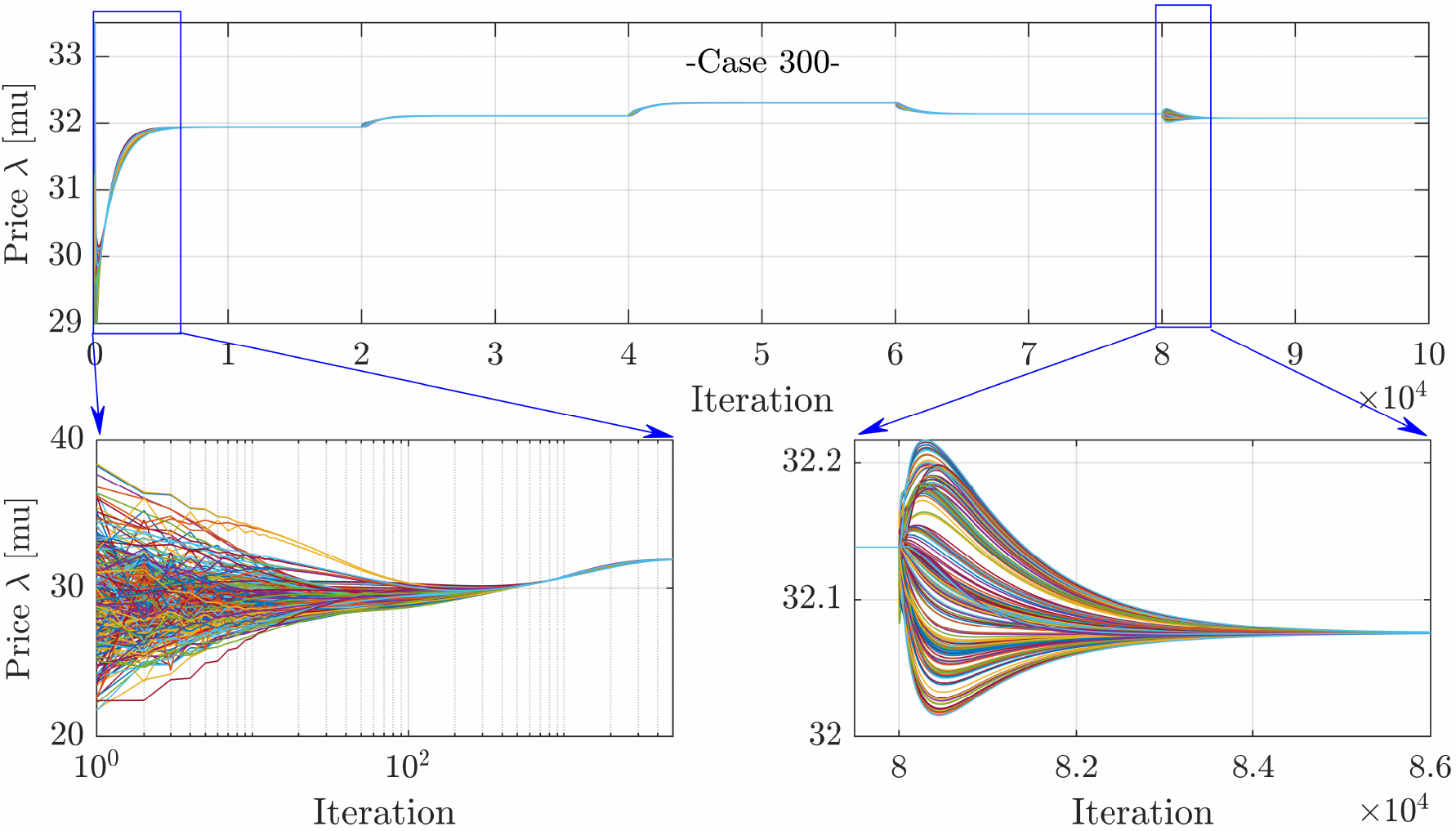}
        \end{minipage}
    \end{minipage}
    \caption{(a) and (d):Consensus on market price for IEEE 30-bus and 300-bus respectively;(b) is magnified version at the beginning for 30-bus and (e) is the plot in a logarithmic scale in order to capture the convergence's dynamic of the algorithm for 300-bus; (c) and (f) in the middle as demand changed for 30-bus and 300-bus respectively}
    \label{fig:300bus_lambda}
\end{figure}

\begin{figure}[h!]
    \begin{minipage}{1\linewidth}
        \begin{minipage}{1\linewidth}
            \includegraphics[width=\linewidth]{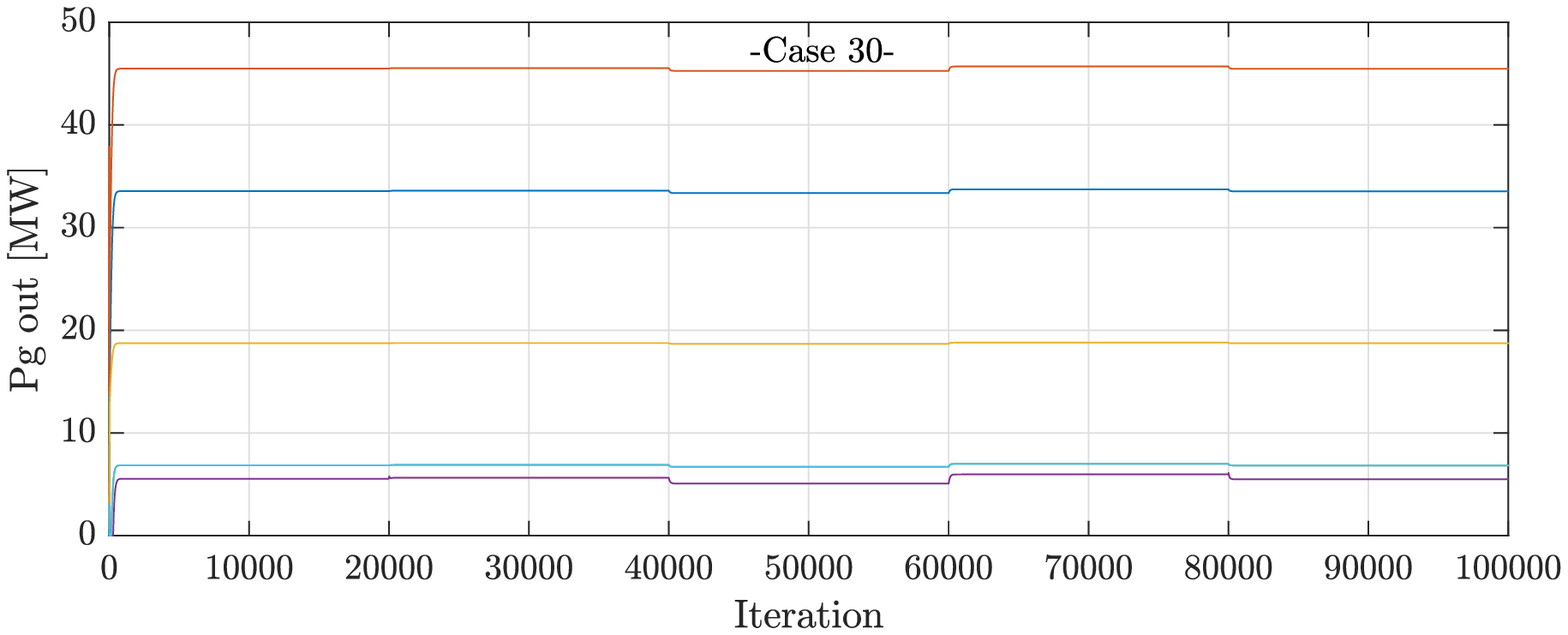}
            \caption{Generators output for  IEEE 30-bus}
            \label{fig:30bus_Pout}
        \end{minipage}
        \vfill
        \begin{minipage}{1\linewidth}
            \includegraphics[width=\linewidth]{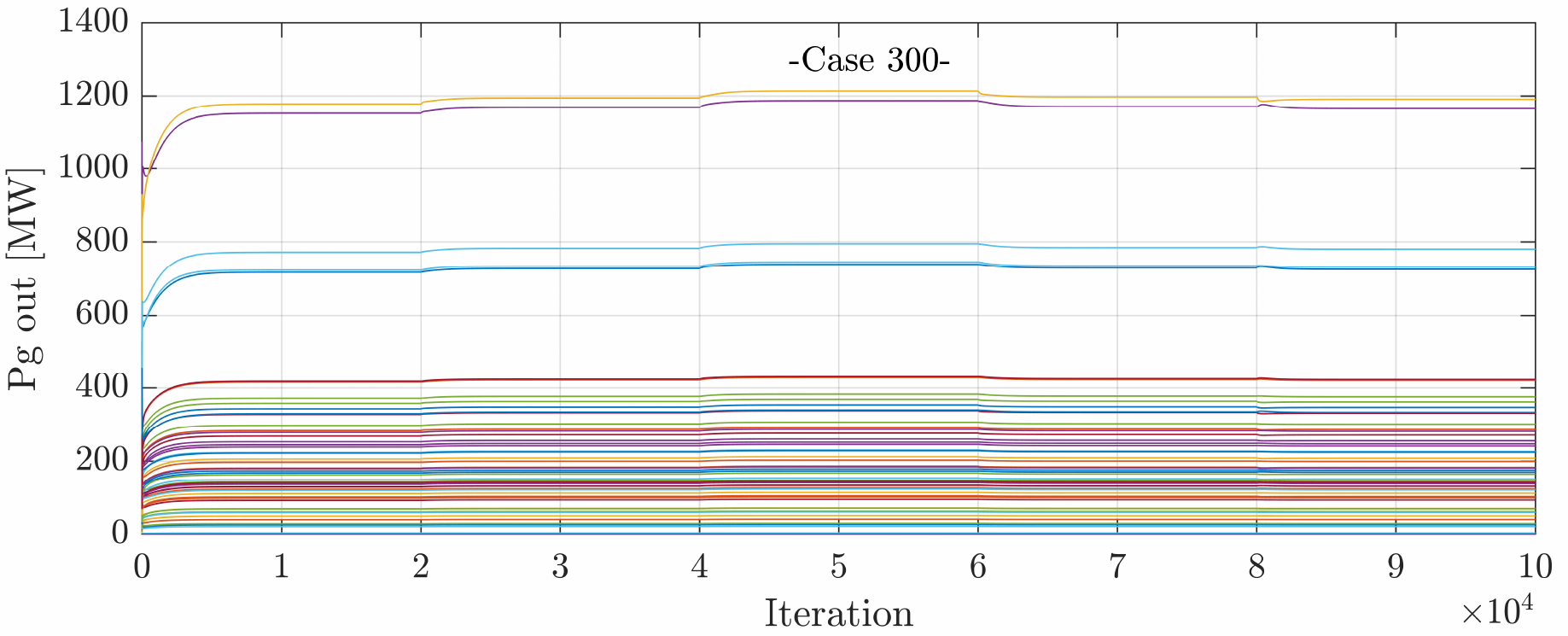}
            \caption{Generators output for IEEE 300-bus}
            \label{fig:300bus_Pout}
        \end{minipage}
    \end{minipage}
\end{figure}

\begin{figure}[h!]
    \begin{minipage}{1\linewidth}
        \begin{minipage}{1\linewidth}
            \includegraphics[width=\linewidth]{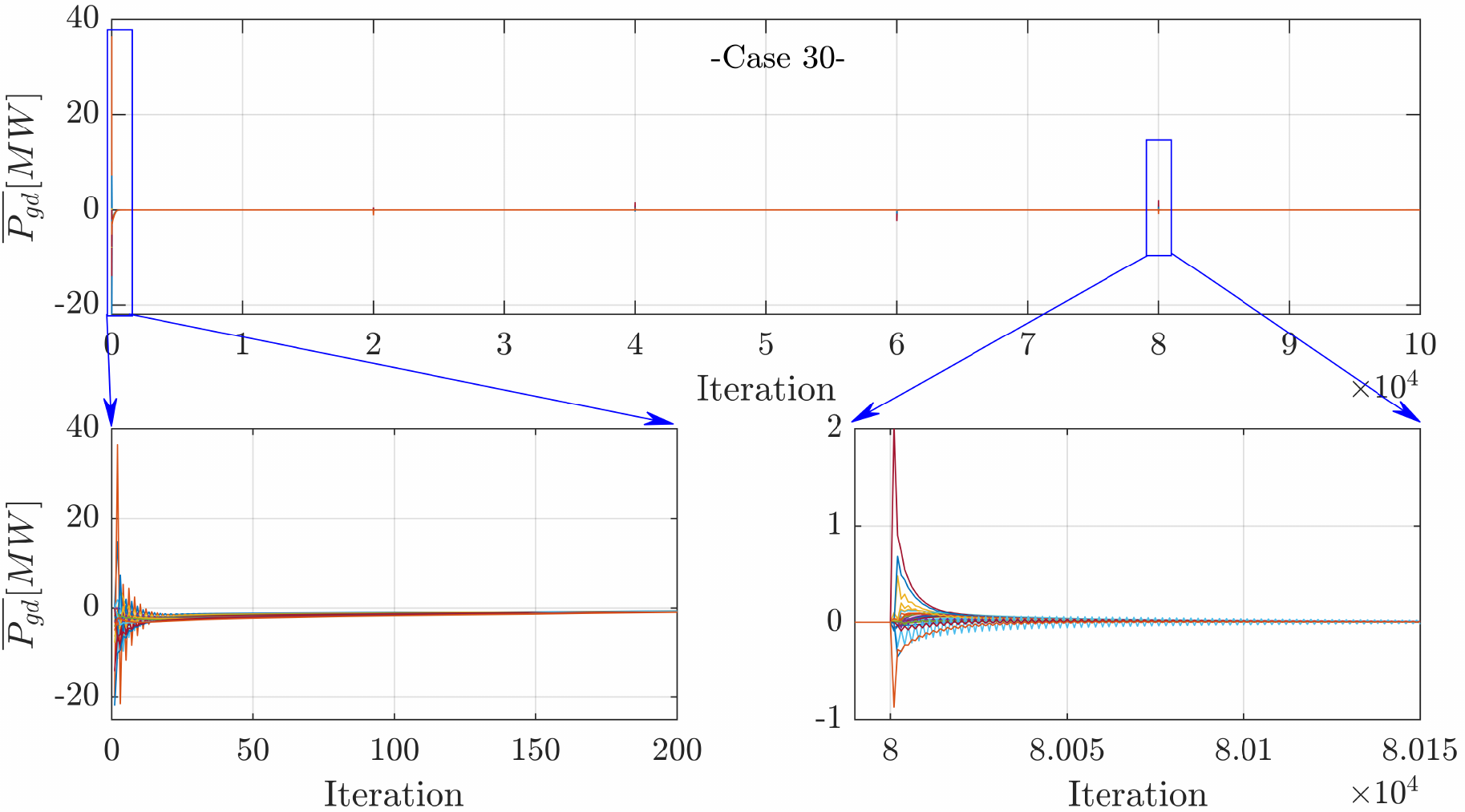}
            \caption{Average power mismatch in IEEE 30-bus}
            \label{fig:30bus_Pgd}
        \end{minipage}
        \vfill
        \begin{minipage}{1\linewidth}
            \includegraphics[width=\linewidth]{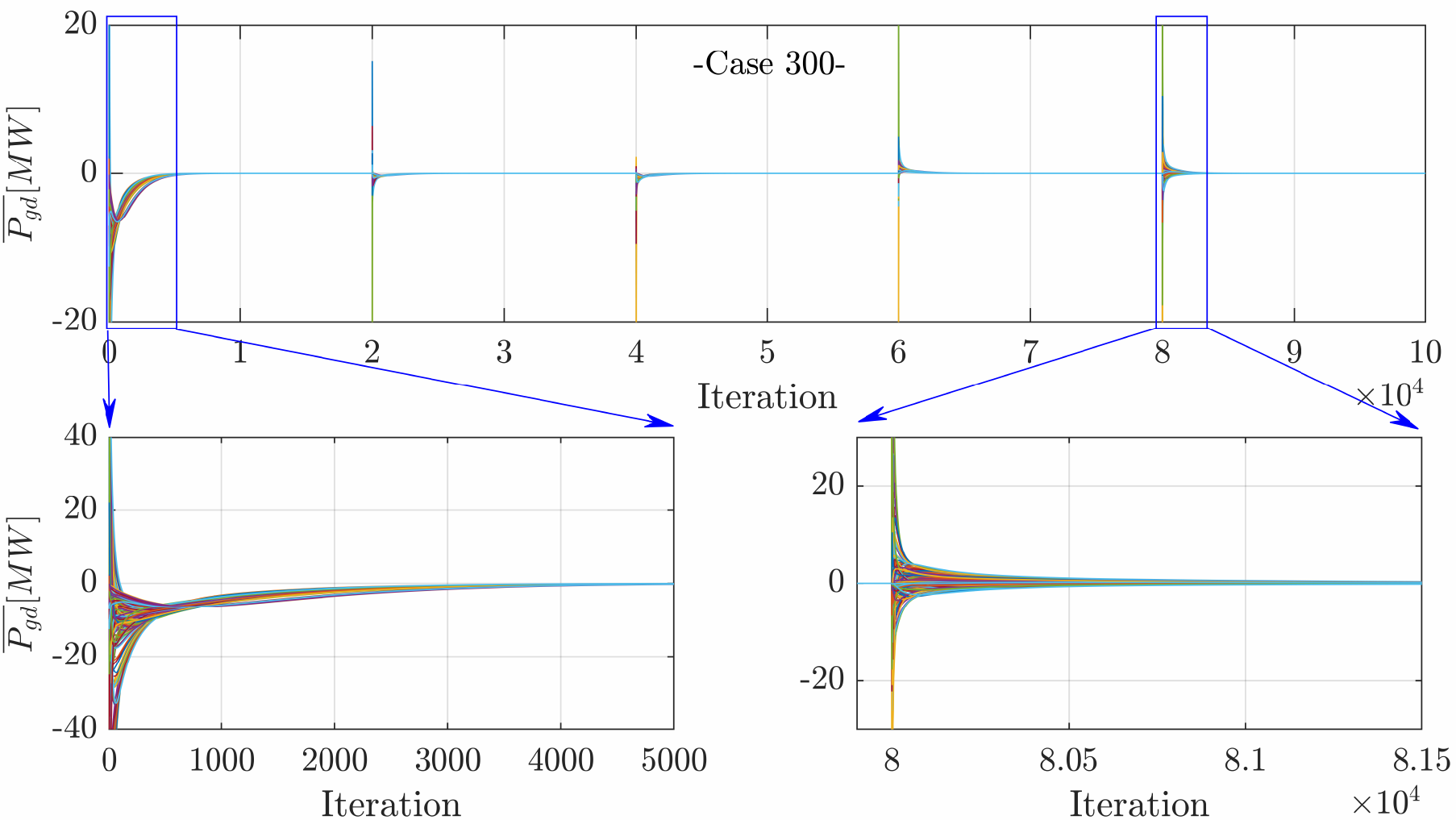}
            \caption{Average power mismatch IEEE 300-bus}
            \label{fig:300bus_Pgd}
        \end{minipage}
    \end{minipage}
\end{figure}

Fig.\ref{fig:300bus_Pgen_Pdem} shows the total generation of the network following the total demand in IEEE 30-bus and 300-bus systems. In IEEE 30-bus case, it takes 600 iterations to catch the demand in the beginning and 400 iterations in subsequent demand change, while it takes 5000 iterations to find the demand in 300-bus system. This validates the claim that the algorithm is robust enough to address the power imbalance. 

Fig. \ref{fig:300bus_lambda} shows the convergence of market energy price $\lambda$ for both 30-bus and 300-bus test cases. The initial consensus of $\lambda$ takes more iterations than in the subsequent demand changes. It is because in the beginning, the agents start from random prices and no consensus exists, while they are at consensus before any following demand changes. Table \ref{table: lambda convergence} details the number of iterations and the computation time elapsed to reach consensus. 

Figs. \ref{fig:30bus_Pout} and \ref{fig:300bus_Pout} depict the output of generators at the optimal point for both cases. These results demonstrate that the total power generation perfectly follows the total demand changes. Since in the proposed algorithm, agents share their own power with their neighbors instead of individual information of power generation and demand to better preserve their privacy, it is important to look at the power mismatch signals and their convergence. 

Figs. \ref{fig:30bus_Pgd} and \ref{fig:300bus_Pgd} depict the agents' estimate of the average power mismatch between generation and demand for 30-bus and 300-bus test systems, respectively. The results illustrate that the agents estimates reach consensus and converge to zero in less than 50 iterations in 30-bus case and less than 80 iterations in 300-bus case. This demonstrates that the agents fulfill the power balance constraint without communicating their detailed information of generation and demand with their neighbors. 

\subsubsection{Computation Time}
 The computer specification used to implement the algorithm is Desktop PC with Intel Core i7 processor (3.6GHz) 64 GB RAM. Table \ref{table: lambda convergence} lists the time taken for the convergence of $\lambda$. The convergence takes in millisecond range for $IEEE 30$ buses to $IEEE 300$ buses, which shows the real -time applicability of the algorithm.  

\section{Conclusion} 
\label{sec:conclusion}
In this paper, the ADMM algorithm is fully distributed to solve the economic dispatch problem, removing the dual updater from the optimization problem. The proposed solution exploits dual decomposition and \textit{dynamic average consensus} algorithms to develop the update procedures with minimal information shared just between neighbors. Performance of the proposed solution including convergence and computation speed is tested against different IEEE test cases at different scales. Simulations demonstrate promising results for the algorithm to solve real-time economic dispatch problems with penetration of solar photo-voltaic.
\section{Appendix}
\subsection{Conservation Property}
\label{sec:Appndx-Cosv prop} 
Rewriting  \eqref{eq:avg_consensus},
\begin{align}
    \mathbf{x}_{}^{k+1} = \mathbf{Ax}_{}^{k}
\end{align}
Adding $\Delta \mathbf{z}$ to our original equation, we obtain:
\begin{align}
    \mathbf{x}_{}^{k+1} = \mathbf{Ax}^{k} +\Delta \mathbf{z}
    \label{eq:plus_delta_z}
\end{align}
If $\mathbf{x}(k)$ is subtracted from both sides of \eqref{eq:plus_delta_z}, we have:
\begin{align}
    \mathbf{x}_{}^{k+1}-\mathbf{x}_{}^{k} = \mathbf{Ax}_{}^{k}+\Delta \mathbf{z}-\mathbf{x}_{}^{k}\\
    \mathbf{1}^T(\mathbf{x}_{}^{k+1}-\mathbf{x}_{}^{k}) = \mathbf{1}^T(\mathbf{Ax}_{}^{k}+\Delta \mathbf{z}-\mathbf{x}_{}^{k})
\end{align}

From the property of matrix \textbf{A}  given by equation (\ref{Eq:necessary cond}), we can write:
\shahab{}{There is an error here}
\begin{align}
    \sum \Delta \mathbf{x} =\mathbf{1}^T\mathbf{x}^k-\mathbf{1}^T\mathbf{x}^k+\mathbf{1}^T\Delta\mathbf{z}\\
    \sum \Delta \mathbf{x}=\sum \Delta \mathbf{z}
    \label{eq:conservation prop}
\end{align} 

Thus, the conservation property at iteration is proven.

\bibliographystyle{IEEEtran}
{\footnotesize \bibliography{IEEEabrv, D_Ed}}

\begin{thebibliography}{10}
\providecommand{\url}[1]{#1}
\csname url@samestyle\endcsname
\providecommand{\newblock}{\relax}
\providecommand{\bibinfo}[2]{#2}
\providecommand{\BIBentrySTDinterwordspacing}{\spaceskip=0pt\relax}
\providecommand{\BIBentryALTinterwordstretchfactor}{4}
\providecommand{\BIBentryALTinterwordspacing}{\spaceskip=\fontdimen2\font plus
\BIBentryALTinterwordstretchfactor\fontdimen3\font minus
  \fontdimen4\font\relax}
\providecommand{\BIBforeignlanguage}[2]{{%
\expandafter\ifx\csname l@#1\endcsname\relax
\typeout{** WARNING: IEEEtran.bst: No hyphenation pattern has been}%
\typeout{** loaded for the language `#1'. Using the pattern for}%
\typeout{** the default language instead.}%
\else
\language=\csname l@#1\endcsname
\fi
#2}}
\providecommand{\BIBdecl}{\relax}
\BIBdecl

\bibitem{kargarian2016toward}
A.~Kargarian, J.~Mohammadi, J.~Guo, S.~Chakrabarti, M.~Barati, G.~Hug, S.~Kar,
  and R.~Baldick, ``Toward distributed/decentralized dc optimal power flow
  implementation in future electric power systems,'' \emph{IEEE Transactions on
  Smart Grid}, vol.~9, no.~4, pp. 2574--2594, 2016.

\bibitem{boyd2011distributed}
S.~Boyd, N.~Parikh, E.~Chu, B.~Peleato, J.~Eckstein \emph{et~al.},
  ``Distributed optimization and statistical learning via the alternating
  direction method of multipliers,'' \emph{Foundations and
  Trends{\textregistered} in Machine learning}, vol.~3, no.~1, pp. 1--122,
  2011.

\bibitem{wang2015dynamic}
T.~Wang, D.~O’Neill, and H.~Kamath, ``Dynamic control and optimization of
  distributed energy resources in a microgrid,'' \emph{IEEE transactions on
  smart grid}, vol.~6, no.~6, pp. 2884--2894, 2015.

\bibitem{disfani2015distributed}
V.~R. Disfani, L.~Fan, and Z.~Miao, ``Distributed dc optimal power flow for
  radial networks through partial primal dual algorithm,'' in \emph{2015 IEEE
  Power \& Energy Society General Meeting}.\hskip 1em plus 0.5em minus
  0.4em\relax IEEE, 2015, pp. 1--5.

\bibitem{xu2013distributed}
Y.~Xu, W.~Zhang, W.~Liu, X.~Wang, F.~Ferrese, C.~Zang, and H.~Yu, ``Distributed
  subgradient-based coordination of multiple renewable generators in a
  microgrid,'' \emph{IEEE Transactions on Power Systems}, vol.~29, no.~1, pp.
  23--33, 2013.

\bibitem{chen2017admm}
G.~Chen and Q.~Yang, ``An admm-based distributed algorithm for economic
  dispatch in islanded microgrids,'' \emph{IEEE Transactions on Industrial
  Informatics}, vol.~14, no.~9, pp. 3892--3903, 2017.

\bibitem{spanos2005dynamic}
D.~P. Spanos, R.~Olfati-Saber, and R.~M. Murray, ``Dynamic consensus on mobile
  networks,'' in \emph{IFAC world congress}.\hskip 1em plus 0.5em minus
  0.4em\relax Citeseer, 2005, pp. 1--6.

\bibitem{spanos2005distributed}
------, ``Distributed sensor fusion using dynamic consensus,'' in \emph{IFAC
  World Congress}.\hskip 1em plus 0.5em minus 0.4em\relax Citeseer, 2005.

\bibitem{newman2018networks}
M.~Newman, \emph{Networks}.\hskip 1em plus 0.5em minus 0.4em\relax Oxford
  university press, 2018.

\bibitem{xiao2004fast}
L.~Xiao and S.~Boyd, ``Fast linear iterations for distributed averaging,''
  \emph{Systems \& Control Letters}, vol.~53, no.~1, pp. 65--78, 2004.

\bibitem{xu2011novel}
Y.~Xu and W.~Liu, ``Novel multiagent based load restoration algorithm for
  microgrids,'' \emph{IEEE Transactions on Smart Grid}, vol.~2, no.~1, pp.
  152--161, 2011.

\bibitem{guo2015distributed}
F.~Guo, C.~Wen, J.~Mao, and Y.-D. Song, ``Distributed economic dispatch for
  smart grids with random wind power,'' \emph{IEEE Transactions on Smart Grid},
  vol.~7, no.~3, pp. 1572--1583, 2015.

\bibitem{Jussi:2018}
\BIBentryALTinterwordspacing
J.~Ahola, \emph{DEE Photovoltaic Power Plant Weather Station Data}, 2018
  (accessed Nov 3, 2019). [Online]. Available: \url{http://www.tut.fi/solar/}
\BIBentrySTDinterwordspacing

\end{thebibliography}
\end{document}